\documentclass[11pt,letterpaper]{article}

\usepackage{graphicx}
\usepackage{epstopdf}
\usepackage{subfigure}

\title{Path Lengths in Turbulence}
\author{Nicholas T. Ouellette$^{1}$, Eberhard Bodenschatz$^{2,3,4}$ and Haitao Xu$^{2}$\\
$^1$ Department of Mechanical Engineering \& Materials Science \\
Yale University, New Haven, CT 06520, USA \\
$^2$ Max Planck Institute for Dynamics and Self-Organization \\
D-37077 G{\" o}ttingen, Germany\\
$^3$ Institute for Nonlinear Dynamics, University of G\"ottingen \\
D-37077 G\"ottingen, Germany\\
$^4$ Laboratory of Atomic and Solid State Physics and \\
Sibley School of Mechanical and Aerospace Engineering\\
Cornell University, Ithaca, NY 14853, USA
}

\date{\today}

\begin{document}

\maketitle

\begin{abstract}
By tracking tracer particles at high speeds and for long times, we study the geometric statistics of Lagrangian trajectories in an intensely turbulent laboratory flow. In particular, we consider the distinction between the displacement of particles from their initial positions and the total distance they travel. The difference of these two quantities shows power-law scaling in the inertial range. By comparing them with simulations of a chaotic but non-turbulent flow and a Lagrangian Stochastic model, we suggest that our results are a signature of turbulence.
\end{abstract}

\section{Introduction}

Understanding turbulent flow remains a challenge for both theory and experiment. With strongly coupled active degrees of freedom spanning orders of magnitude in space and time, intense turbulence has so far resisted a complete description. Although dimensional models have provided valuable guidance \cite{K41}, we still lack a fully predictive theory of turbulence. 

Recent years, however, have seen the exciting development of new tools---both experimental and numerical---for measuring the Lagrangian trajectories of fluid elements in turbulence \cite{toschi2009}. This capability has in turn spurred the formulation of theoretical ideas that take advantage of the newly available information \cite{falkovich2001}. In particular, a broad class of geometric models that explicitly consider the Lagrangian dynamics of the flow has arisen. Many of these models describe the shape of clusters of fluid elements \cite{chertkov1999,biferale2005,luthi2007,xu2008}, in an extension of classical theories of turbulent relative dispersion \cite{richardson1926,batchelor1950,ott2000,bourgoin2006,ouellette2006c}. The geometric properties of \emph{individual} trajectories, however, have also been recently been studied, viewing the trajectories as three-dimensional space curves and focusing on their curvature \cite{braun2006,xu2007,choi2010}.

Here, we study individual particle trajectories from a different standpoint, blending elements of turbulent transport with geometric models. Going back to the seminal work of Taylor \cite{taylor1922}, single-particle turbulent transport theories have described the evolution of $\mathbf{R}(t)$, the vector giving the displacement of a particle from its position at some reference time. The mean-squared displacement $\langle R^2(t) \rangle$ (where $R$ is the modulus of $\mathbf{R}$) is expected to scale as $t^2$ for short times and then as $t$ for $t \gg T_L$, where $T_L$ is the integral time scale \cite{pope2000}, similar to molecular diffusion processes, though with a flow-dependent coefficient. Displacement, however, is not the only distance measure we can apply to a particle path; since we follow individual particles in three dimensions, we can also study the arc length of the trajectories, \textit{i.e.}, the total distance travelled by individual particles. We denote this arc length by $S(t)$. $S$ and $R$ are identical for simple unidirectional flow; in highly rotational flow, however, as is the case in turbulence, we expect significant differences between the two. Since coherent structures in turbulence are expected to be strongly vortical, the difference between $S$ and $R$ may possibly mark these structures.

We find that although the statistics of $S$ alone are similar to those of $R$, the difference of the two shows very different behavior. We measured both $\langle S^2 - R^2 \rangle$ and $\langle (S-R)^2 \rangle$, which differ by the non-negligible covariance $\langle SR \rangle$. Each of these differences shows power-law behavior in the inertial range, though with different exponents; we find that $\langle S^2 - R^2 \rangle \sim t^3$ (similar to classical Richardson dispersion for particle pairs) and $\langle (S-R)^2 \rangle \sim t^{3.7\pm0.2}$. In an effort to isolate the contributions of turbulence to our findings (rather than simple kinematic effects \cite{merrifield2010}), we also measured these statistics in two synthetic flows: the Arnold--Beltrami--Childress (ABC) flow \cite{dombre1986}, which shows chaotic particle motion but is not turbulent, and Sawford's second-order Lagrangian stochastic model \cite{sawford1991}, which approximates non-intermittent turbulent transport but does not have spatial structure. Comparing our experimental results with these synthetic flow fields, we conjecture that a large separation of time scales, and therefore a (Lagrangian) inertial range, is responsible for the detailed results we find.

In Sec.~\ref{sec:experiment} below, we briefly describe our experimental flow and the measurement techniques we used to gather our data. We then present and discuss the statistics in Sec.~\ref{sec:data}, and finally summarize our findings in Sec.~\ref{sec:conclusion}.

\section{Experimental Details}
\label{sec:experiment}

To generate fully developed turbulence in the laboratory, we used a von K{\' a}rm{\' a}n swirling flow of water between counter-rotating baffled disks. Details of the apparatus are given elsewhere \cite{ouellette2006c}. By rotating the disks at frequencies up to 5 Hz, we drove turbulence at Taylor-scale Reynolds numbers $R_\lambda = \sqrt{15 u' L / \nu}$, where $u'$ is the root-mean-square velocity, $L$ is the integral length scale, and $\nu$ is the kinematic viscosity, up to 815. Here, we show data for $R_\lambda = 690$ and 815. Cooling water circulated through the top and bottom plates of the apparatus in order to maintain a constant temperature in the working fluid.

We measured the turbulence by tracking the motion of small, solid tracer particles seeded in the flow. These polystyrene particles have a density of 1.06 g/cm$^3$ and a diameter of 25 $\mu$m, and have previously been shown to be passive tracers in this flow \cite{voth2002}. With RMS speeds of order 1 m/s and accelerations of order $10^2$ m/s$^2$, very fast cameras are required to record the trajectories accurately. We imaged the particles at rates of up to 27 000 frames per second. We used three cameras arranged around the mid-plane of the tank in order to measure the three-dimensional positions of the particles by stereomatching the individual images. In order to study the long-time transport of the particles, we illuminated a measurement volume of $5\times 5 \times 5$ cm$^3$ with two pulsed, frequency-doubled Nd:YAG lasers delivering roughly 150 W of laser light. For comparison, the integral length scale $L$ is 7.1 cm in this flow.

We used particle-tracking methods to make quantitative measurements. We process the raw movies with a multi-frame predictive tracking algorithm \cite{ouellette2006a} that reconstructs the long-time trajectories of individual particles. In order to keep the stereomatching ambiguities at a manageable level, the seeding density of tracers is kept low (roughly 100 per image). Even so, particles can occasionally drop out of view, leading to gaps in the raw trajectories. In a post-processing step, we bridge these gaps by re-tracking the data in a six-dimensional space spanned by the particle positions and velocities \cite{xu2008b}. In this way, we obtain longer tracks that give us access to later-time statistics. Velocities and accelerations are subsequently computed by convolving the measured trajectories with a smoothing and differentiating kernel \cite{mordant2004}.

\section{Results and Discussion}
\label{sec:data}
\subsection{Distance and Displacement}

\begin{figure}
\centering
\includegraphics[width=0.75\linewidth]{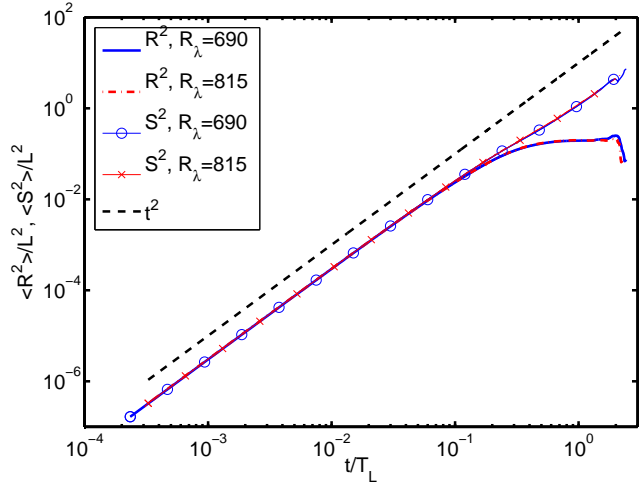}
\caption{\label{fig:RSsq} Evolution of the mean-squared displacement $\langle R^2\rangle$ and distance travelled $\langle S^2 \rangle$ as a function of time for two different Reynolds numbers. Both scale ballistically ($\sim t^2$) at short times, and trend downwards towards diffusive scaling ($\sim t$) at longer times. The dashed line is a $t^2$ power law as a guide for the eye.}
\end{figure}

The simplest transport quantity that can be computed for a single fluid element is its mean-squared displacement from some initial position. This quantity, which we denote as $\langle R^2 \rangle$, is simple to model in turbulence \cite{taylor1922,pope2000,davidson2004}: we expect it to scale as $u'^2 t^2$ for $t \ll T_L$, where $T_L$ is the integral time scale, and as $u'^2 T_L t$ for $t \gg T_L$. Our measurements of $\langle R^2 \rangle$ are shown in Fig.~\ref{fig:RSsq}, and show no unusual behavior.

Displacement, however, is not the only way to measure the change in position of a fluid element. Since we know its full trajectory, we can also measure the total distance travelled in a time $t$ by measuring the arc length of the trajectory. We label this positive-definite quantity $S$, and study its mean-square value $\langle S^2 \rangle$, as shown in Fig.~\ref{fig:RSsq}. Like $\langle R^2 \rangle$, we find that $\langle S^2 \rangle \sim t^2$ for short times, with nearly the same scaling constant. This behavior should be expected: at short times, distance and displacement are very similar. 

\subsection{Difference Statistics}

More intriguing than the statistics of $\langle R^2 \rangle$ and $\langle S^2 \rangle$ alone are the statistics of their \emph{difference}, for two main reasons. First, since both $\langle R^2 \rangle$ and $\langle S^2 \rangle$ scale like $t^2$ for $t \ll T_L$, we may be able to observe sub-leading contributions to their scaling by subtracting them. Additionally, large discrepancies between the displacement of a fluid element and the distance it has travelled may be associated with highly vortical regions, and may be indicative of coherent structures such as vortex tubes. 

\begin{figure}
\centering
\subfigure[]{
\includegraphics[width=0.47\linewidth]{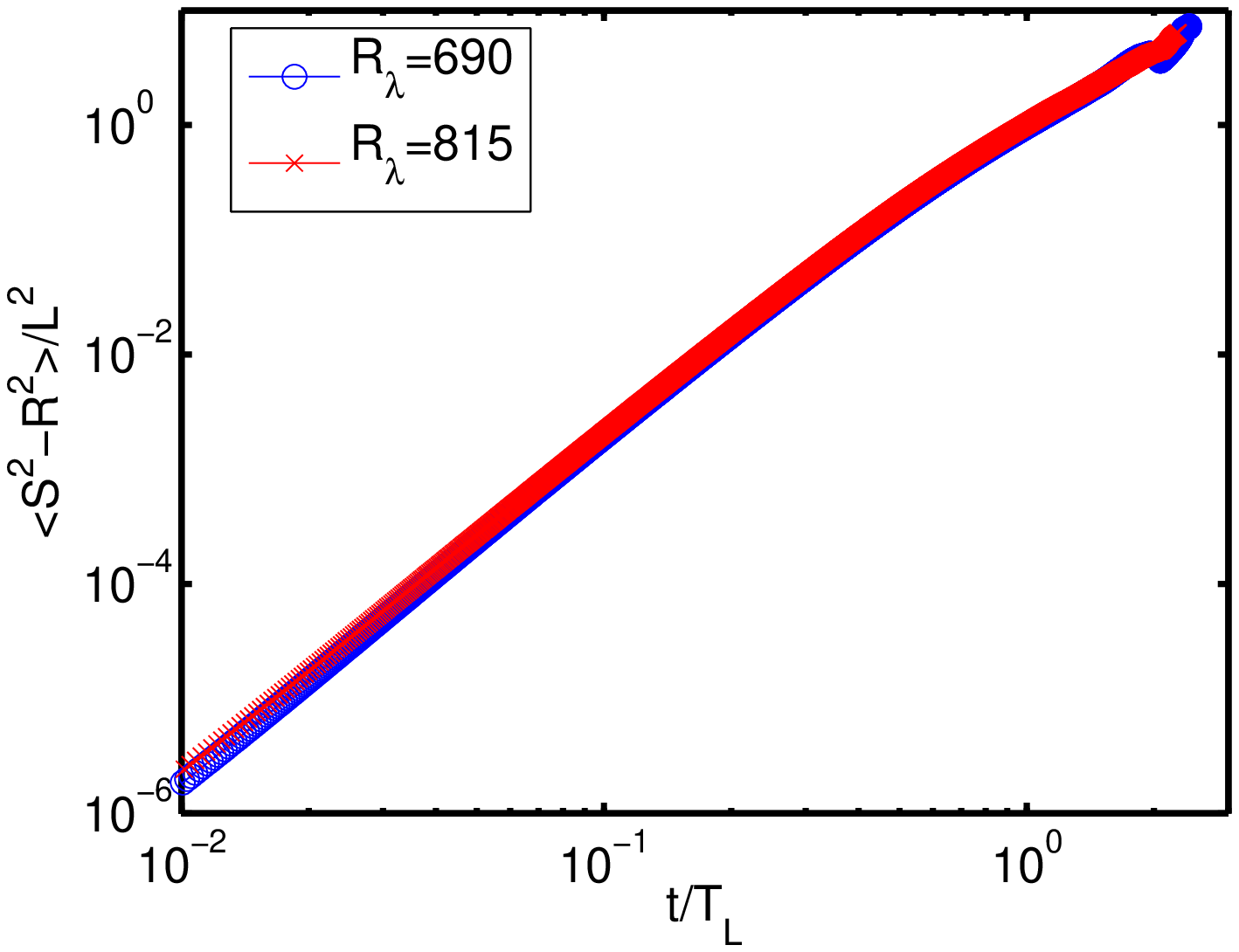}
\label{fig:Ssq_Rsq}
}
\subfigure[]{
\includegraphics[width=0.47\linewidth]{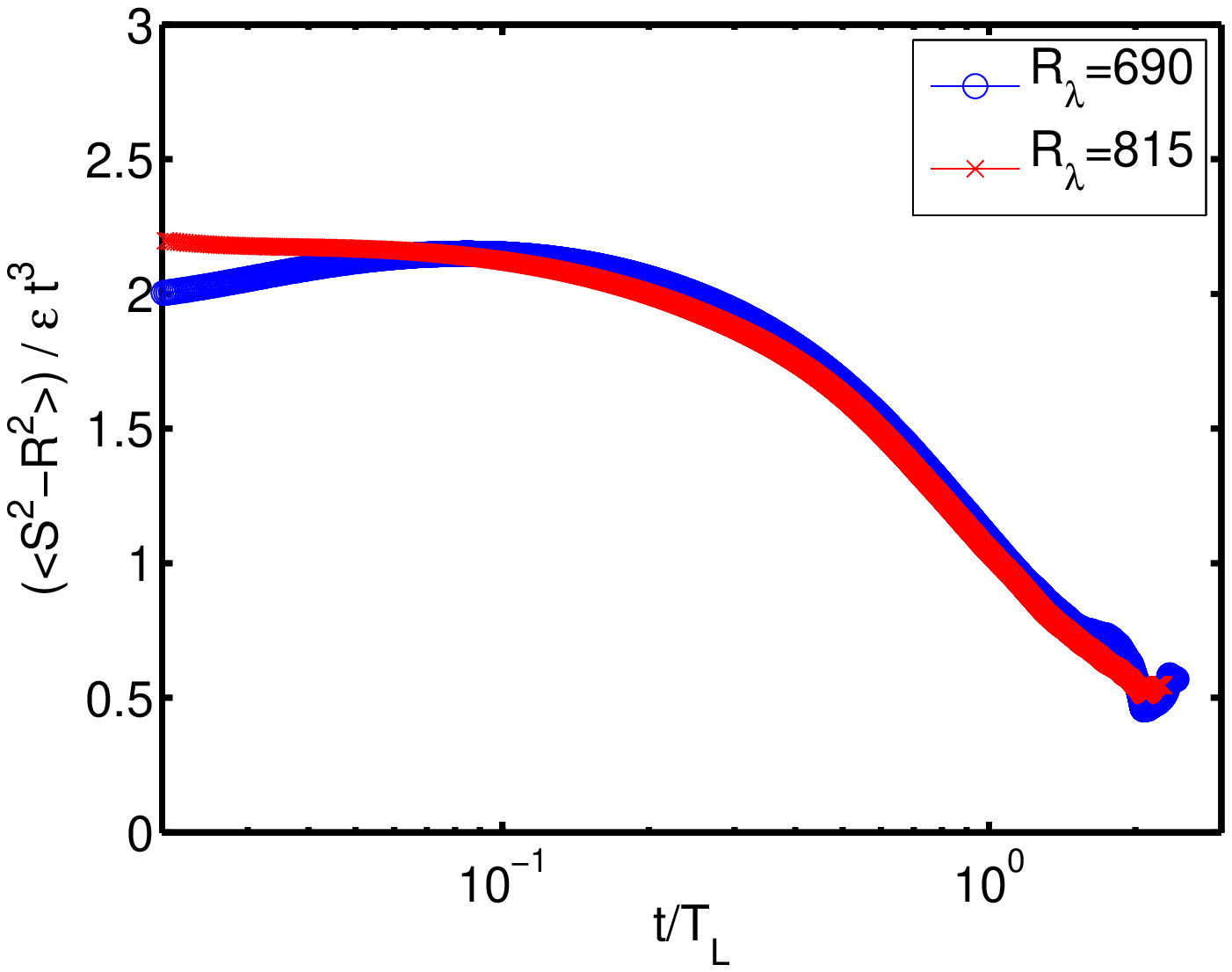}
\label{fig:Ssq_Rsq_compensated}
}
\caption{\label{fig:Ssq_Rsq_all} The difference $\langle S^2 - R^2 \rangle$ as a function of time, plotted in (a) logarithmic coordinates and (b) compensated by $\epsilon t^3$, where $\epsilon$ is the energy dissipation rate. In the inertial range, this difference appears to scale as $\epsilon t^3$ with a scaling constant of roughly 2 and a scaling region that grows with Reynolds number.}
\end{figure}

\begin{figure}
\centering
\includegraphics[width=0.66\linewidth]{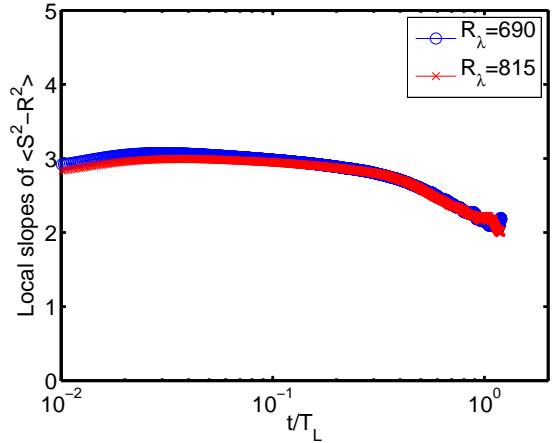}
\caption{\label{fig:ls_Ssq_Rsq} Local slopes (i.e., the logarithmic derivative) of $\langle S^2 - R^2 \rangle$ as a function of time, confirming the $t^3$ scaling in the inertial range.}
\end{figure}

In Fig.~\ref{fig:Ssq_Rsq}, we show the simple difference $\langle S^2 \rangle - \langle R^2 \rangle$ as a function of time. At the two Reynolds numbers shown, this difference scales as a power law in the inertial range. Unlike $\langle R^2 \rangle$ and $\langle S^2 \rangle$ alone, however, this difference scales as $t^3$ rather than $t^2$, similar to the prediction for Richardson scaling of the mean-squared separation of particle \emph{pairs} in turbulence. This observed scaling can be rationalized by applying Kolmogorov-style reasoning: $\langle S^2 \rangle - \langle R^2 \rangle$ has dimensions of length squared, and thus must scale like $\epsilon t^3$ (where $\epsilon$ is the typical energy dissipation rate per unit mass) in the inertial range. We check this scaling prediction by compensating $\langle S^2 \rangle - \langle R^2 \rangle$ by $\epsilon t^3$, as shown in Fig.~\ref{fig:Ssq_Rsq_compensated}. To check the robustness of this $t^3$ power law, we performed a local-slope analysis by computing the logarithmic derivative
\begin{equation}
\frac{\mathrm{d} \log \langle S^2 - R^2 \rangle}{\mathrm{d} \log t},
\end{equation}
which gives the local scaling exponent of the data as a function of time. Robust power laws will have long plateaus in a local-slopes plot. Our results are shown in Fig.~\ref{fig:ls_Ssq_Rsq}. Local slopes have been shown to be a clean way of identifying power laws in experimental data sets without the need to fit to potentially small domains \cite{biferale2008,arneodo2008}. This analysis confirms the existence of a $t^3$ regime. The $t^3$ regime therefore appears to be a distinctive feature of high-Reynolds-number turbulence.

\begin{figure}
\centering
\subfigure[]{
\includegraphics[width=0.47\linewidth]{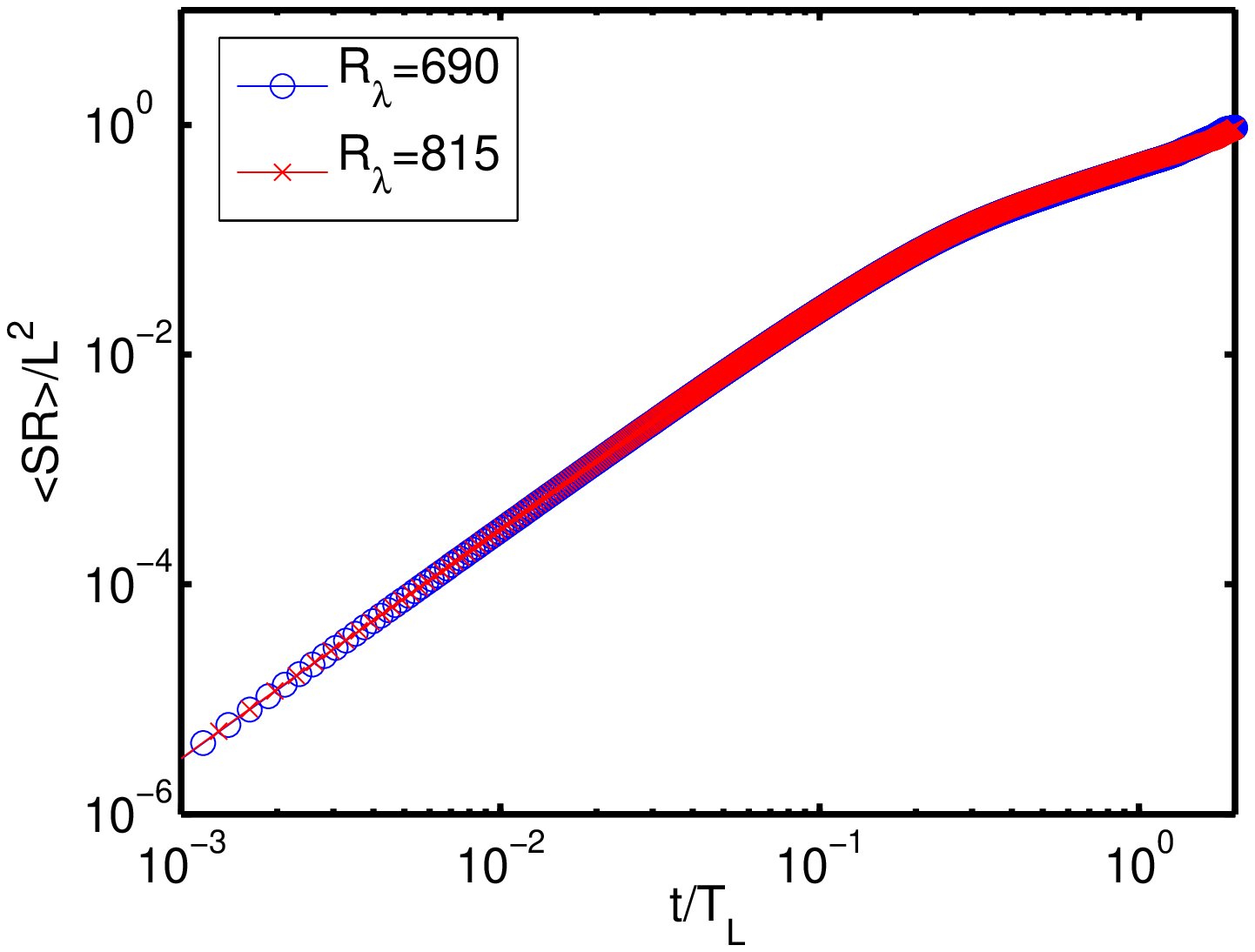}
\label{fig:SR}
}
\subfigure[]{
\includegraphics[width=0.47\linewidth]{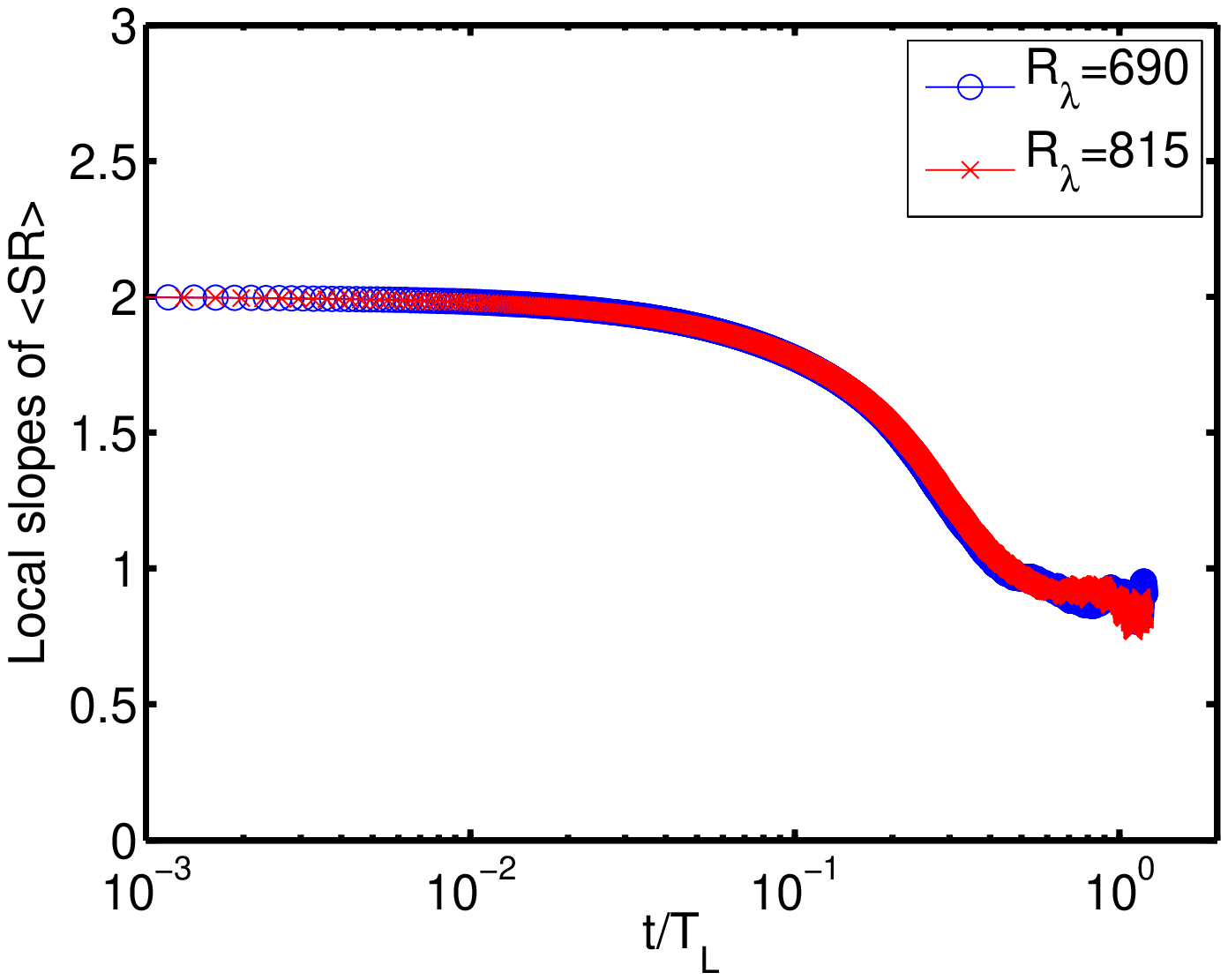}
\label{fig:ls_SR}
}
\caption{\label{fig:SR_all} (a) The covariance $\langle SR \rangle$ plotted in logarithmic coordinates. Just as for $\langle S^2 \rangle$ and $\langle R^2 \rangle$, $\langle SR \rangle$ scales as $t^2$ at short times and subsequently as $t$ at longer times. (b) Local slopes for $\langle SR \rangle$, confirming the scaling laws.}
\end{figure}

But even though these data show a scaling regime compatible with Kolmogorov's 1941 theory \cite{K41} (commonly known as K41), it is not clear that K41 can be applied. 
In K41, scaling predictions can only be made for scale-local quantities.
Differences can often be assumed to be scale-local \cite{batchelor1950}, but this heuristic argument cannot distinguish between $\langle S^2 \rangle - \langle R^2 \rangle$ and $\langle (S-R)^2 \rangle$, two quantities that are dimensionally equivalent.
The distinction between these two statistics is the covariance $\langle SR \rangle$; cross terms such as this one have been shown to play an important role in obtaining clean scaling in turbulent relative dispersion \cite{bourgoin2006,ouellette2006c}. In Fig.~\ref{fig:SR_all}, we show our measurements of $\langle SR \rangle$. As expected, this covariance is not at all negligible (since $S$ and $R$ are highly correlated), and grows as a power law in time. As shown by a local-slope analysis in Fig.~\ref{fig:ls_SR}, $\langle SR \rangle$ scales as $t^2$ at short times and $t$ at longer times, just like $\langle S^2 \rangle$ or $\langle R^2 \rangle$. 

\begin{figure}
\centering
\subfigure[]{
\includegraphics[width=0.47\linewidth]{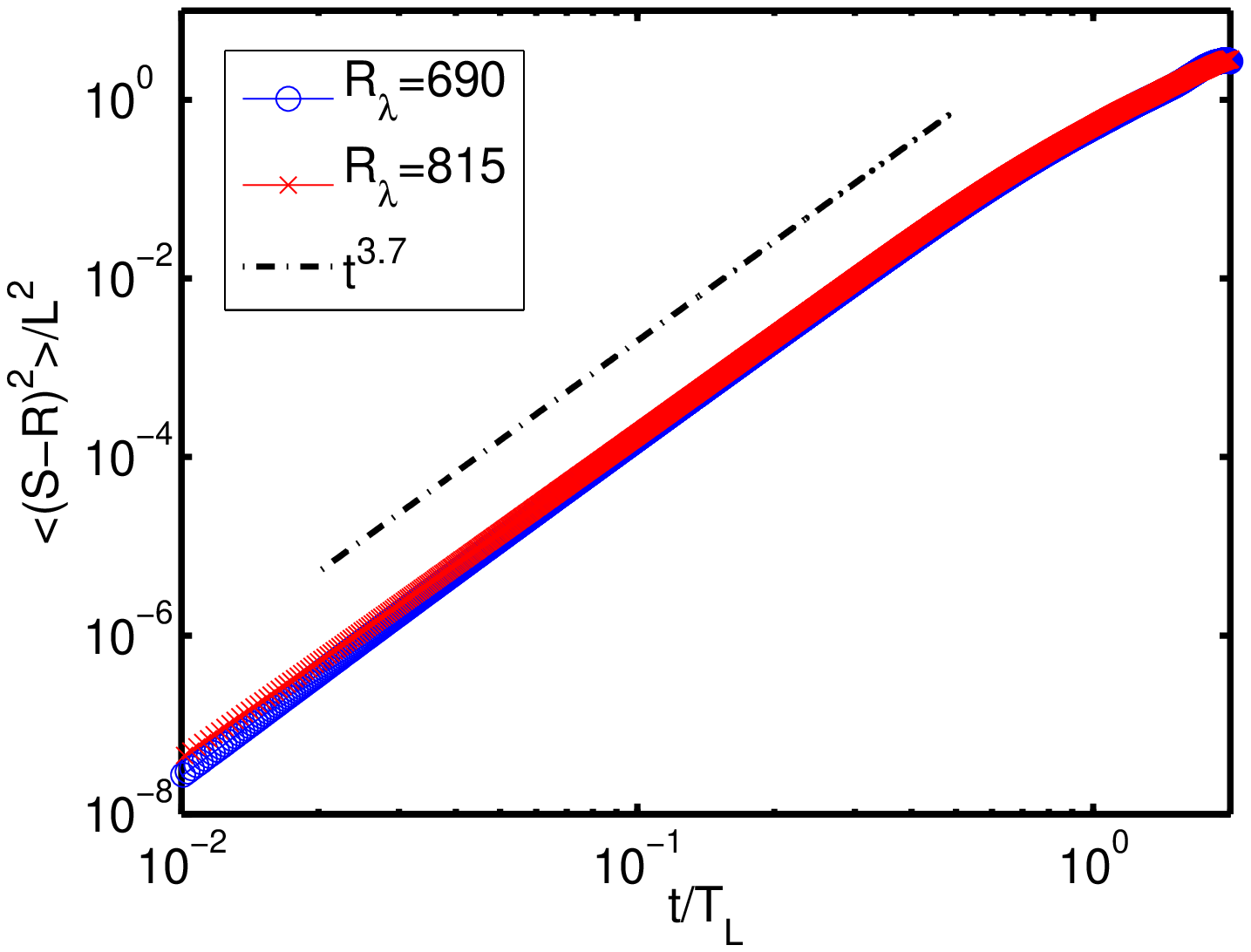}
\label{fig:SRdiffSq}
}
\subfigure[]{
\includegraphics[width=0.47\linewidth]{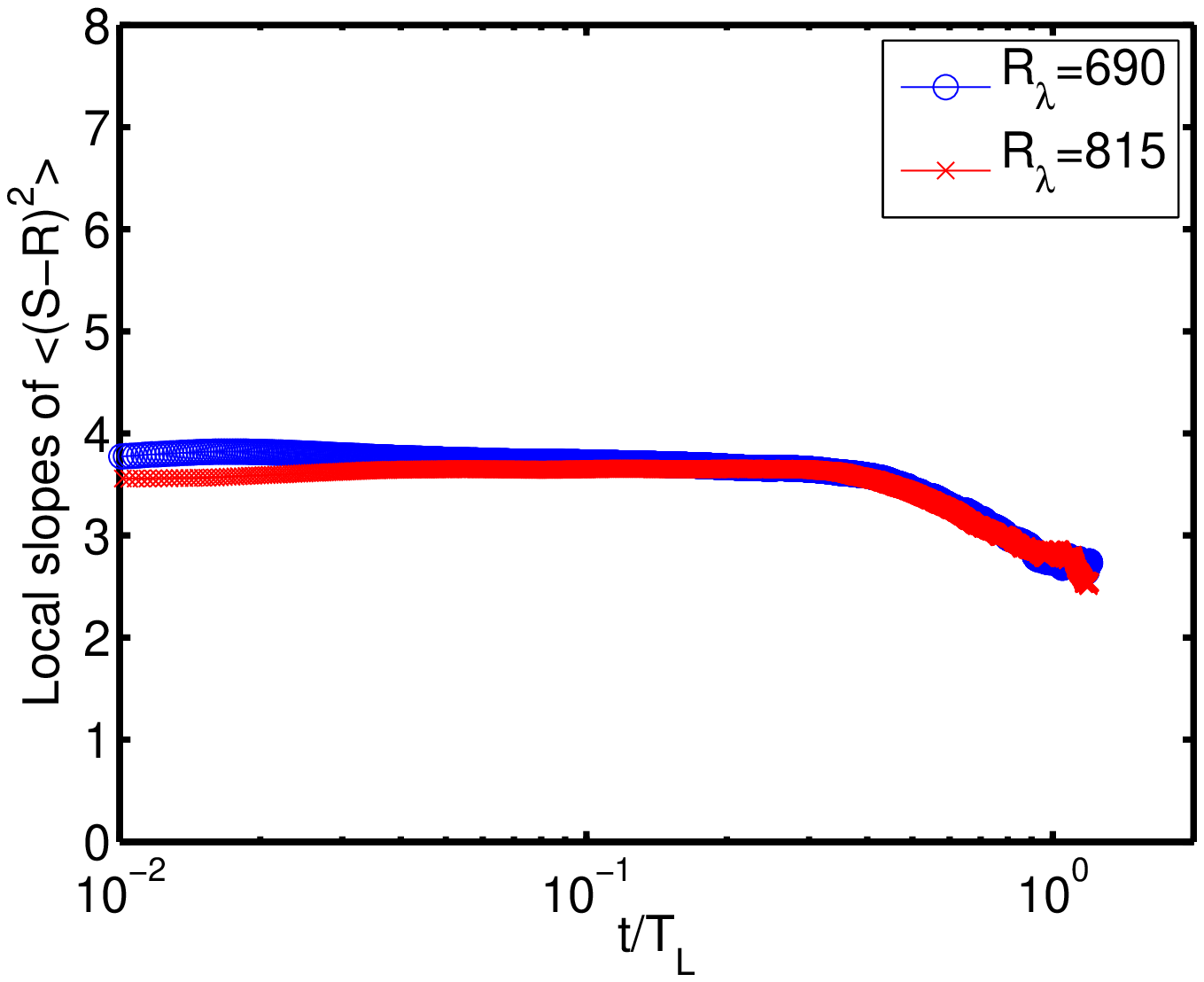}
\label{fig:ls_SRdiffSq}
}
\caption{\label{fig:SRdiffSq_all} (a) The difference $\langle(S-R)^2\rangle$ plotted in logarithmic coordinates. The straight line is a $t^{3.7}$ power law as a guide for the eye.
(b) Local slopes for $\langle(S-R)^2\rangle$, suggesting that the inertial-range $t^{3.7}$ scaling is robust.}
\end{figure}

We therefore consider the quantity $\langle (S-R)^2 \rangle$, which we show in Fig.~\ref{fig:SRdiffSq_all}. For inertial-range times it scales as $t^{3.7\pm0.2}$, which is seen clearly from a local-slope analysis, shown in Fig.~\ref{fig:ls_SRdiffSq}. The local-slope analysis also suggests that (as expected) the temporal extent of the inertial-range power law scaling region grows with $R_\lambda$. A scaling exponent of $3.7$ cannot be obtained from K41; and since these statistics are only second order, any possible intermittency corrections are likely to be small. As we confirm below, this $t^{3.7}$ regime appears to be unique to turbulence and requires the existence of an extended inertial range, but does not need intermittent particle accelerations or velocity increments.

\subsection{Comparison with Other Flows}

In order to ensure that the statistical signatures we see are truly related to the turbulence and not simply to kinematics \cite{merrifield2010}, we also simulated particle trajectories in a chaotic but non-turbulent flow and in a Lagrangian stochastic model. 

\subsubsection{ABC Flow}

The steady Arnold--Beltrami--Childress (ABC) flow is a canonical example of a flow field that has a simple spatial structure but  can show complex, chaotic Lagrangian dynamics \cite{dombre1986}. The velocity field, which is a solution of the Euler equations, is given by
\begin{eqnarray}
\mathbf{u}(x,y,z) = &\left( A\sin kz + C\cos ky \right)\hat{\mathbf{e}}_x \nonumber \\
&+ \left( B\sin kx + A\cos kz \right)\hat{\mathbf{e}}_y \nonumber \\
&+ \left( C\sin ky + B\cos kx \right) \hat{\mathbf{e}}_z,
\end{eqnarray}
where $A$, $B$, and $C$ are constants, $k$ is the spatial wavenumber of the flow, and $\hat{\mathbf{e}}_x$, $\hat{\mathbf{e}}_y$, and $\hat{\mathbf{e}}_z$ are unit vectors in the three Cartesian directions. Even though the spatially periodic flow field is steady, fluid elements advected in this flow will decorrelate on a time scale related to the time necessary to exit a unit cell. We estimate this time scale to be 
\begin{equation}
T_{ABC} = \frac{2\pi/k}{\left(A^2 + B^2 + C^2\right)^{1/2}}.
\end{equation}

\begin{figure}
\centering
\includegraphics[width=0.66\linewidth]{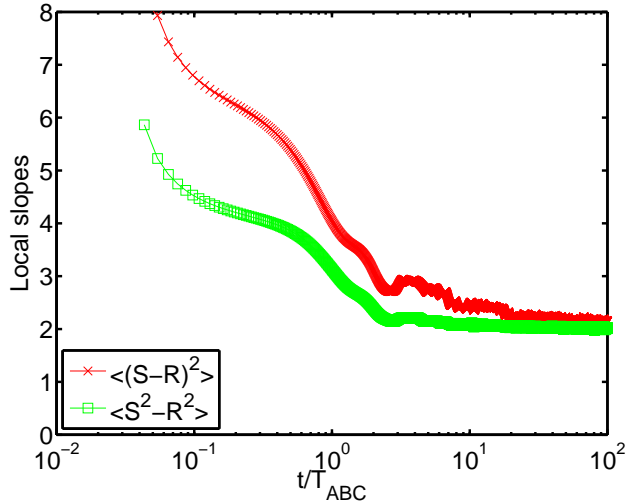}
\caption{\label{fig:ABC} Local slopes for $\langle S^2 - R^2 \rangle$ and $\langle (S-R)^2 \rangle$ as measured in the ABC flow. The inertial-range scalings seen for the turbulent flow are not present.}
\end{figure}

We simulated a population of fluid elements evolving in the ABC flow using a fourth-order Runge-Kutta algorithm, choosing $A=12$, $B=15$, and $C=10$ \cite{sapsis2008}, and measured $S(t)$ and $R(t)$ along their trajectories. In Fig.~\ref{fig:ABC}, we show the local slopes for $\langle (S-R)^2\rangle$ and $\langle S^2 - R^2 \rangle$ for these fluid elements. At long times, both quantities saturate to a $t^2$ scaling. This result is to be expected since for $t \gg T_{ABC}$, $\langle R^2(t) \rangle \sim t$ and $\langle S^2(t) \rangle \sim t^2$; at long times, the leading-order $t^2$ term dominates. For shorter times, we observe no clean power-law scaling. Each curve in Fig.~\ref{fig:ABC} has a short region that may indicate a short power-law regime; the exponents, however, are very different from the $t^3$ and $t^{3.7}$ scalings seen in the turbulence data. 
The ABC flow misses some of the ingredients of turbulence.

\subsubsection{Lagrangian Stochastic Model}

Although our measurements in the ABC flow do not show the same scaling as the turbulence measurements for $\langle S^2 - R^2 \rangle$ and $\langle (S-R)^2 \rangle$, it remains to be seen whether our observations are a result of the broad spectral activity in turbulence (that is, the presence of both a large and a small time scale separated by an inertial range) or are due to intermittency. We therefore used Sawford's second-order Lagrangian stochastic model \cite{sawford1991} to test these possibilities.

Sawford's model assumes that the material derivative of the acceleration is a white-noise process, so that the Lagrangian acceleration is given by a stochastic differential equation. The velocity and position of a fluid element are obtained by successive integrations of the acceleration. Though this model has two time scales (corresponding to the integral time $T_L$ and the Kolmogorov time $\tau_\eta$) and therefore has an inertial range, it is not intermittent: the probability density functions of the acceleration and the velocity increments are Gaussian.

\begin{figure}
\centering
\includegraphics[width=0.66\linewidth]{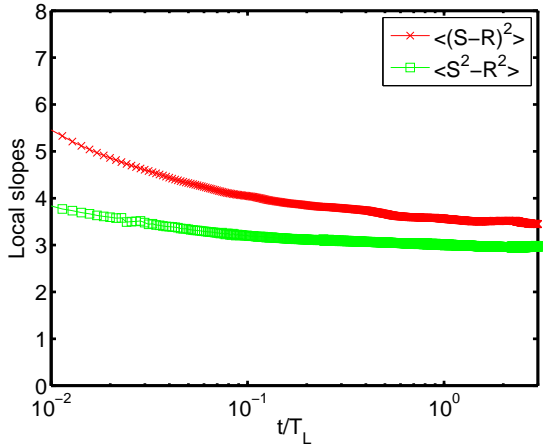}
\caption{\label{fig:sawford} Local slopes for $\langle S^2 - R^2 \rangle$ and $\langle (S-R)^2 \rangle$ computed using Sawford's model \cite{sawford1991}. The observed turbulent inertial-range scaling is evident here, despite the absence of intermittency in the model.}
\end{figure}

Our results computed using this model with $R_\lambda = 815$ (averaged over many realizations) are shown in Fig.~\ref{fig:sawford}. We see behavior much more similar to the turbulence than to the ABC flow, though with shorter scaling ranges at similar Reynolds numbers. These results suggest that we are observing normal inertial range scaling rather than the anomalous scaling associated with intermittency, a result that is to be expected for low-order statistical moments like the ones we consider here.

\section{Summary}
\label{sec:conclusion}

We measured statistics of the trajectories of single particles in an intensely turbulent laboratory flow, focusing on the displacement of the particles from their initial positions and the total distance travelled by the particles. The mean-squared values of each of these lengths scales as expected, growing ballistically at short times and diffusively at long times. Differences of the distance and displacement, however, reveal new scaling laws that depend on exactly how the difference is taken. Although we do not have a detailed theoretical explanation for our observations at this time, we conjecture that our results are a signature of the separation of time scales present in turbulent flow, and may possibly be signatures of Lagrangian structures in the flow. We hope that these results stimulate theoretical investigations of the measured statistics.

\section*{acknowledgements}
N.T.O. acknowledges support from the U.S. National Science Foundation under Grant No.~DMR-0906245. H.X. and E.B. are supported by the Max Planck Society. We thank the Kavli Institute for Theoretical Physics (supported by the National Science Foundation under Grant No.~PHY-0551164) for hospitality and support during the writing of this paper.

\end{document}